\documentclass[prl,showpacs,twocolumn,longbibliography]{revtex4-1}

\usepackage{xr}
\usepackage{hyperref}

\usepackage{color}
\usepackage[usenames,dvipsnames]{xcolor}
\usepackage{amsmath,amsthm,amssymb}
\usepackage{graphicx}
\usepackage{epsfig}
\usepackage{dcolumn}
\usepackage{bm}
\usepackage{mathrsfs}
\usepackage{multirow}
\usepackage[all]{xy}
\usepackage{pbox}
\usepackage{verbatim}

\DeclareMathOperator{\Tr}{Tr}

\def\(({\left(}
\def\)){\right)}
\def\[[{\left[}
\def\]]{\right]}

\newcommand{\be}{\begin{equation}}
\newcommand{\ee}{\end{equation}}
\newcommand{\ben}{\begin{eqnarray}}
\newcommand{\een}{\end{eqnarray}}
\newcommand{\beq}{\begin{equation}}
\newcommand{\eeq}{\end{equation}}

\newcommand{\la}{\langle}
\newcommand{\ra}{\rangle}

\newcommand{\e}{{\text{e}}}

\begin{document}

\title{Current fluctuations in boundary-driven quantum spin chains}

\author{Federico Carollo, Juan P. Garrahan, and Igor Lesanovsky}
\affiliation{School of Physics and Astronomy}
\affiliation{Centre for the Mathematics and Theoretical Physics of Quantum Non-Equilibrium Systems,
University of Nottingham, Nottingham, NG7 2RD, UK}

\date{\today}

\begin{abstract}
Boundary-driven spin chains are paradigmatic non-equilibrium systems in both classical and quantum settings. In general it may not be possible to distinguish classical from quantum transport through monitoring the mean current, as both ballistic as well as diffusive regimes occur in either setting. Here we show that genuine quantum features become manifest in large fluctuations which allow a discrimination between classical and quantum transport: in the classical case, realizations that are characterized by atypically large boundary activity are associated with larger than typical currents, i.e. an enhanced number of events at the boundaries goes together with a large current. Conversely, in the quantum case the Zeno effect leads to the suppression of current in trajectories with large activity at the boundary. We analyze how these different dynamical regimes are reflected in the structure of rare fluctuations. We show furthermore that realizations supporting a large current are generated via weak long-range correlations within the spin chain, typically associated with hyperuniformity.
\end{abstract}

\maketitle
\begin{figure}[h!]
\centering
\includegraphics[scale=0.30]{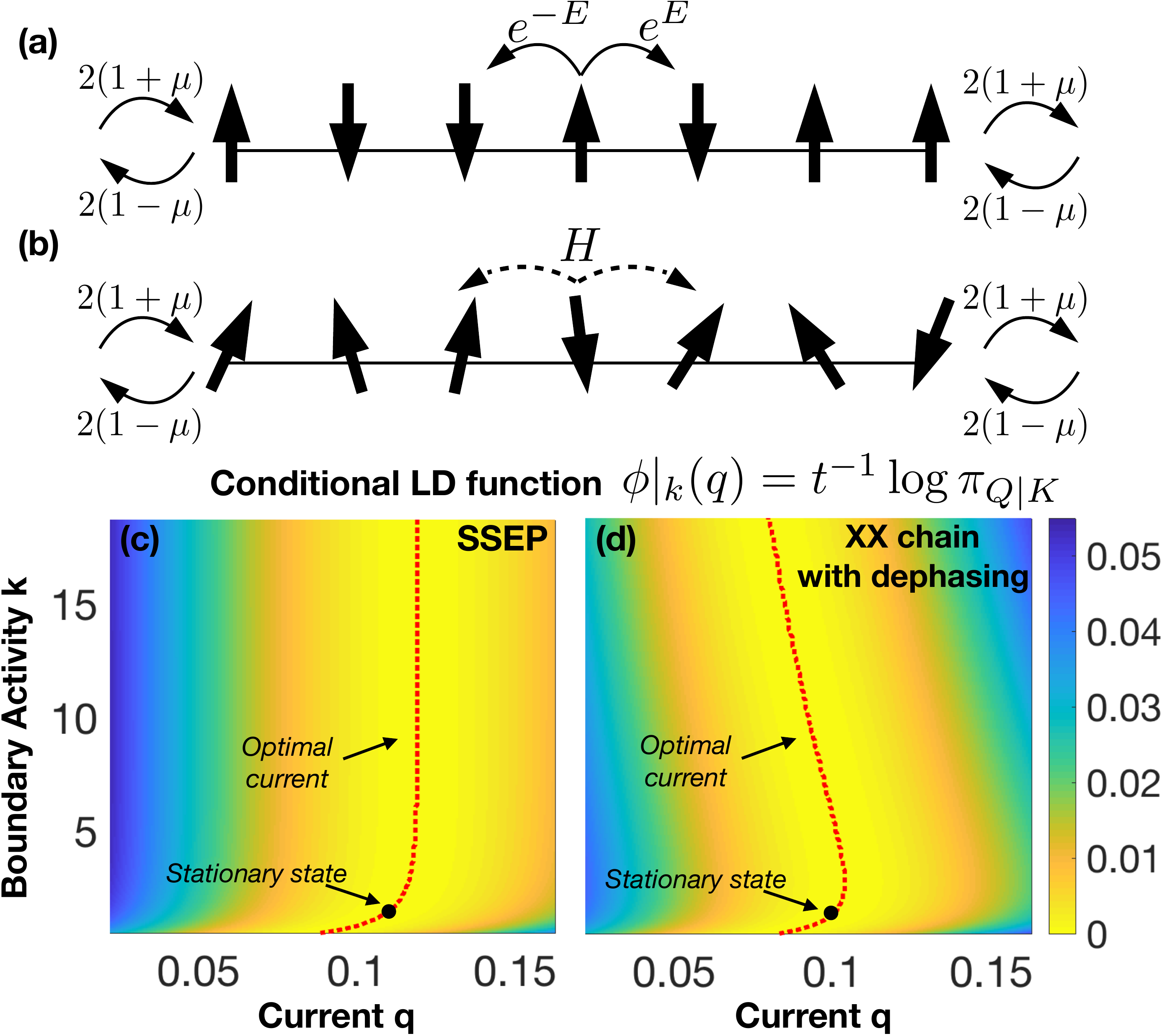}
\caption{{\bf (a)} Classical chain: an up arrow indicates the presence of a particle. The latter jumps into neighboring sites only if these are empty, and rates can be asymmetric in the presence of a field $E\neq0$. Particle injection/ejection takes place with boundary driving parameter $\mu\in[0,1]$. {\bf (b)} Quantum chain: transport is governed by an Hamiltonian $H$. Arrows with different angles indicate superposition states. {\bf (c)}-{\bf (d)} Conditional large deviation (LD) function $\phi|_{k}(q)$ for a system of $L=6$ sites, $\mu=0.6$. The dashed line represents the most likely observed value of the current $q$, for different  activities $k$, while bullets indicate the stationary state behavior.  {\bf(c)} SSEP with $E=0$. {\bf(d)} XX-chain with dephasing showing suppression of the current for large activities. }
\label{fig11}
\end{figure}

 
\noindent {\bf \em Introduction.} 
Complex collective behavior of non-equilibrium systems admits a simplified thermodynamic-like description in terms of few macroscopic degrees of freedom \cite{Touchette2009,Chetrite2015,Bertini2015}. These are usually time-averaged observables accounting for the response of the system to some quench or to some external driving \cite{Bodineau2004,Derrida2007,Derrida2009,Hurtado2014,Bernard2016,Castro-Alvaredo2016,Karevski2017,Baek2017,Ljubotina:2017aa}. For boundary-driven chains [\emph{cf.} Fig.~\ref{fig11}(a)-(b)], a complete characterization of the collective dynamics is achieved by considering the integrated current $Q(t)$ -  the net number of particles leaving the system through one of the boundaries up to time $t$ - and the activity $K(t)$ - the total number of jumps at the same boundary. Typically, one is interested in average values of these quantities; however, interesting collective phenomena are often to be found in rare dynamical realizations \cite{PhysRevE.78.021122,G2009first,Garrahan2010,Espigares2013,Jack2015,Baek2017}. 
Moreover, the majority of studies on non-equilibrium systems focus on properties of the stationary state, see e.g. \cite{ZnidXXSstate,Znidaric2010b,Prosen2011,ILIEVSKI2014485,PhysRevLett.101.105701,PhysRevB.94.054313,Karevski2017bb}, and only a minority analyze rare dynamical behaviors \cite{Garrahan2010,Znidaric2014,Znidaric2014b,PhysRevB.90.115156,Manzano2017}, mainly focused on the statistics of currents. In particular, much progress has been made in predicting the non-equilibrium behavior of different quantum and classical systems, unifying them in few classes according to the macroscopic emergent type of particle transport (e.g.~diffusive, ballistic). 
In specific cases, quantum systems have even been mapped to classical counterparts. For instance, it has been shown that the current statistics of the XX-chain in the presence of dephasing coincides, in the thermodynamic limit, to the one of the classical symmetric exclusion process (SSEP) \cite{PhysRevE.96.052118,Znidaric2014b}. The question that naturally arises from these results, is thus whether there exist universal features in the rare events of quantum non-equilibrium systems distinguishing these from classical ones, or whether, at this macroscopic level, the microscopic quantum nature of the system becomes irrelevant. 

In this work, we show that there is indeed one marked difference between classical and quantum boundary driven transport and, irrespectively of microscopic details of the dynamics and of the emergent collective behavior, a clear distinction between  quantum and classical regime can be established at the level of large fluctuations. We quantify the properties of dynamical fluctuations through the probability $\pi_{K,Q}$ of a dynamical realization with activity $K$ and current $Q$. For long times $t$, this probability obeys a so-called large deviation (LD) principle \cite{Touchette2009,Chetrite2015}, i.e. $\pi_{K,Q}\approx \e^{-t\, \phi(k,q)}$, where $k(t)=t^{-1}K(t)$ and $q(t)=t^{-1}Q(t)$. The function $\phi(k,q)$ is positive, and becomes zero when both its arguments $k$ and $q$ take the stationary state values $\la k\ra$ and $\la q\ra$.  Similarly, the conditional probability of a current $Q$, given an activity $K$ is $\pi_{Q|K}\approx e^{-t\phi|_{k}(q)}$, with $\phi|_{k}(q)=\phi(k,q)-\phi(k)$, and $\phi(k)$ being the LD function of the activity. In Fig.~\ref{fig11}(c)-(d), we show this conditional LD function for paradigmatic classical (SSEP) and quantum (XX-chain with dephasing) spin chains, where the difference becomes obvious: for the classical case, increasing activity leads to an increase in currents, as seen in the increase of the optimal current - i.e. the most likely observed current for given activity. In contrast, in the quantum case, large activities lead to a suppression of currents, as seen by the decrease of the optimal current.

In the following we consider in detail both quantum boundary-driven spin chains (XX-chain and XXZ-chain) and classical ones (exclusion processes). For these models, we discuss the full range of fluctuations focusing, in particular, on the interplay between current and activity. Using perturbative arguments we moreover establish that the features displayed in Fig.~\ref{fig11} are universal, i.e. they hold for any quantum spin chain Hamiltonian. Finally, we discuss how the distinct fluctuation behavior in the classical and quantum transport regime manifests in the spatial structure of particle trajectories. \\

\noindent {\bf \em Spin chain models and dynamical large deviation formalism.} We briefly introduce the details of the formalism needed to derive the statistical properties of events mediated by the boundary driving. In order to obtain the joint current-activity statistics, it is convenient to work with the moment generating function $\mathcal{Z}_{s,h}=\sum_{K,Q}\e^{-sK+hQ}\pi_{K,Q}$. This function not only provides all moments of the observables, but it can also be interpreted as the \emph{dynamical partition function} of an ensembles of biased probabilities, $\pi^{s,h}_{K,Q}=\e^{-sK+hQ}\pi_{K,Q}$, favoring or disfavoring different realizations according to the value of the outcomes. These ensembles are associated to rare dynamical behaviors of the system, and are often used to describe properties of large fluctuations \cite{PhysRevLett.95.010601,Lecomte2007,G2009first,MPSGarrahan}. Indeed, for long-times $\mathcal{Z}_{s,h}\asymp \e^{\psi(s,h)\,t}$, and $\psi(s,h)$ is the cumulant generating function (CGF) of the time-averaged quantities $k(t),q(t)$ in both typical and biased ensembles of trajectories. As an example, while $\langle q(t)\rangle=\partial_h \psi(s,h)|_{s,h=0}$ is the average current in the steady-state, $\langle q(t)\rangle_{s,h}=\partial_h \psi(s,h)=(t\mathcal{Z}_{s,h})^{-1}\sum_{K,Q}Q\,\pi^{s,h}_{K,Q}$ is the  current in the $s,h$-ensemble of rare trajectories.
The LD function $\phi(k,q)$ can then be obtained via the Legendre transform $\phi(k,q)=\max_{s,h}\left[-sk+hq-\psi(s,h)\right]$. \\

We use this formalism here to study both classical and quantum spin-$1/2$ chains with $L$ sites, and connected through their first and last site to thermal reservoirs [see Fig.~\ref{fig11}(a)-(b)]. For quantum systems, we consider coherent transport due to the Hamiltonian  
\begin{equation}
H=\sum_{k=1}^{L-1}\left(\sigma_{\rm x}^{(k)}\sigma_{\rm x}^{(k+1)}+\sigma_{\rm y}^{(k)}\sigma_{\rm y}^{(k+1)}+\delta_z \, \sigma_{\rm z}^{(k)}\sigma_{\rm z}^{(k+1)}\right)\, ,
\label{H}
\end{equation}
where $\sigma_{\rm \alpha}^{(k)}$, is the $\alpha$ Pauli matrix of the $k$-th spin. The external driving,  describing dissipative particle injection/ejection at the boundary sites, is given by \cite{Lindblad1976,Gorini1976}
\begin{equation}
\begin{split}
&\mathcal{D}_{s,h}[\cdot]=\sum_{\alpha=0}^1\gamma_{\alpha}\left(L_\alpha^{(1)} \cdot L_\alpha^{{(1)}\dagger}-\frac{1}{2}\left\{\cdot,\left(L^\dagger_\alpha L_\alpha\right)^{(1)}\right\}\right)+\\
&\sum_{\alpha=0}^1\gamma_{1-\alpha}\left(\e^{-s-(-1)^{\alpha}h}L_{\alpha}^{(L)} \cdot L_{\alpha}^{{(L)}\dagger}-\frac{1}{2}\left\{\cdot,\left(L^\dagger_{\alpha} L_{\alpha}\right)^{(L)}\right\}\right)\, ;
\label{tgen}
\end{split}
\end{equation}
respectively $L_{0/1}=\sigma_{\pm}$, and $\gamma_{0/1}=2(1\pm\mu)$, with $\mu\in[0,1]$ being the driving parameter. The fields $s$ and $h$, conjugated respectively to the activity and to the current, are used, in the LD formalism, to obtain the CGF $\psi(s,h)$ \cite{Garrahan2010,Znidaric2014b,Znidaric2014,PhysRevB.90.115156,carollo2017}. The latter is given by the eigenvalue with the largest real part of the so-called \emph{tilted operator} $\mathcal{L}_{s,h}[\cdot]=-i[H,\cdot]+\mathcal{D}_{s,h}[\cdot]$.\\

For classical systems, instead, we consider exclusion processes with bulk generator
\begin{equation}
\begin{split}
W_E&=\sum_{k=1}^{L-1}\Big[\e^{E}\sigma_+^{(k+1)}\sigma_-^{(k)} +\e^{-E}\sigma_-^{(k+1)}\sigma_+^{(k)}+\\
&-\e^{E}n^{(k)}\left({\bf 1}_2-n\right)^{(k+1)}-\e^{-E}n^{(k+1)}\left({\bf 1}_2-n\right)^{(k)}\Big]\, ,
\end{split}
\label{ME}
\end{equation}
where $n$ is the number operator, ${\bf 1}_2$ the $2\times2$ identity matrix, and $E$ an external field. Boundaries are accounted for by 
\begin{equation}
\begin{split}
&W^{\rm bound}_{s,h}=\gamma_{0}\left[\sigma_+ -({\bf 1}_2-n)\right]^{(1)}+\gamma_1\left[\sigma_--n\right]^{(1)}+\\
&+\gamma_{1}\left[\e^{-s-h}\sigma_+ -({\bf 1}_2-n)\right]^{(L)}+\gamma_0\left[\e^{-s+h}\sigma_--n\right]^{(L)}\, .
\end{split}
\end{equation}
Also in this case, $W^{\rm tot}_{s,h} = W_E + W^{\rm bound}_{s,h}$ constitutes the tilted operator from which all cumulants of the observables are extracted \cite{Lecomte2007,G2009first}. \\

\noindent {\bf \em Current fluctuations in the XX-chain.} 
\begin{figure}[h!]
\centering
\includegraphics[scale=0.29]{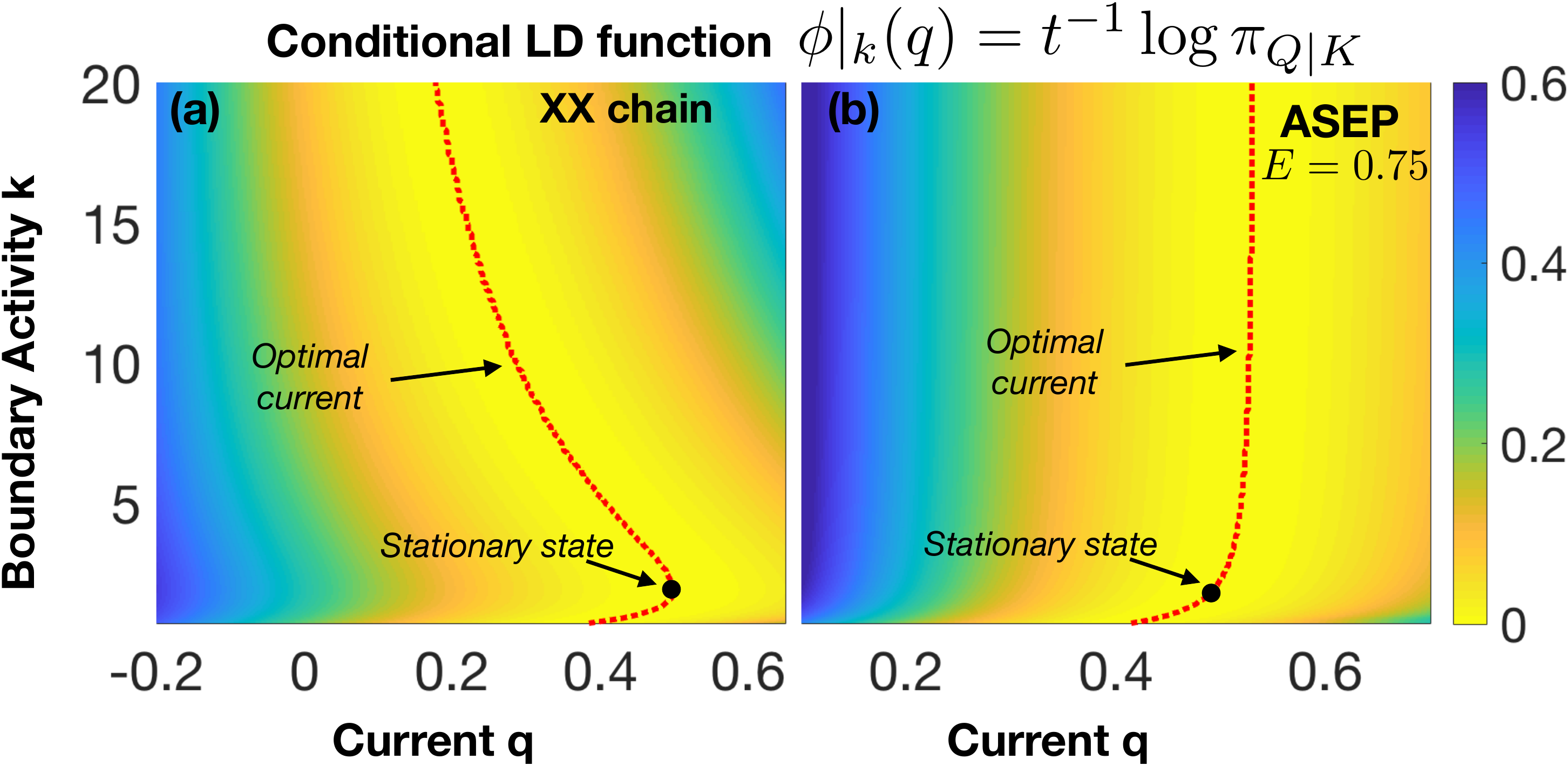}
\caption{Plot of the conditional LD function $\phi|_{k}(q)$: the dashed lines indicate the most likely value of the current, points instead the stationary state behavior. {\bf (a)} XX-chain for $\mu=0.5$. {\bf (b)} Classical ballistic asymmetric exclusion process (ASEP), for $L=6$, $\mu=0.5$ and $E=0.75$.}
\label{fig1}
\end{figure}
We start by considering the XX-chain, whose Hamiltonian is given by Eq.~\eqref{H} with $\delta_z=0$. This is a non-interacting ballistic system. In order to compute $\psi(s,h)$, we exploit that the map $\mathcal{L}_{s,h}[\cdot]$ 
can be cast into the non-Hermitian operator 
\cite{prosen2008third,prosen2010spectral,Znidaric2014} 
\begin{equation}
\hat{\mathcal{L}}_{s,h}={\bf a}\cdot U^\dagger \begin{pmatrix}
X&0\\
0&-X^T
\end{pmatrix}
 U\cdot{\bf a}-4,\quad U=\frac{1}{\sqrt{2}}\begin{pmatrix}
1&-i\\
1&i
\end{pmatrix}\, .
\label{GenTQ}
\end{equation}
Here, ${\bf a}$ is a $4L$-dimensional vector, whose entries ${\bf a}_i$ are fermionic Majorana operators, $\{{\bf a}_i,{\bf a}_j\}=\delta_{i,j}$, and the matrix $X$ contains the details of the tilted dynamics (see supplemental material \cite{SM}).  
The operator can then be brought into a diagonal form $\hat{\mathcal{L}}_{s,h}=2\sum_{m=1}^{2L}\Lambda_m\, b'_m b_m-4$, in terms of normal modes $b_m,b'_{m}$, where $\Lambda_m$ are the eigenvalues of the matrix $X$ \cite{prosen2008third,prosen2010spectral,SM}. The CGF is therefore $\psi(s,h)=2\sum_{m\in\Lambda^{+}}{\rm Re}(\Lambda_m)-4$, with $\Lambda^{+}$ being the set of $m$ for which ${\rm Re}(\Lambda_m)>0$. Associated to this, one has the left and right eigenvectors  $\la {\bf L}_{s,h}|=\langle \tilde{0}|\prod_{m\in\Lambda^{+}}b_{m}$,  $|{\bf R}_{s,h}\rangle=\prod_{m\in\Lambda^{+}}b'_{m}|0\rangle$, with $\la \tilde{0}|$, $|0\ra$ being the left, respectively  right vacuum. 

Using this formalism one can show that, in the large $L$ limit, the magnitude of the current is always bounded by the value $4/\pi$. This was noticed in \cite{Znidaric2014} for the current statistics, but we see here that this bound is present for any activity, even when the latter is atypically large. Moreover, Fig.~\ref{fig1} shows the same suppression of the optimal current for increasing activities that we already discussed in presence of dephasing [Fig. \ref{fig11}-(d)]. This appears to be a signature of the quantum nature of these large fluctuations, independent of the transport regime. \\

\noindent {\bf \em Current fluctuations and the Zeno effect.} We will now investigate to what extent the previous findings generalize, and, in particular, whether they apply also to interacting systems. 
To address this situation, it is convenient to move from the description of events with fixed $k$ and $q$, given by the LD function $\phi(k,q)$, to the one based on biased ensembles of probabilities, where average values are under control \cite{Lecomte2007,G2009first,MPSGarrahan,PhysRevLett.111.120601}. This means that we turn our attention to the CGF $\psi(s,h)$. We further restrict to $\mu\neq1$, since, as displayed in Fig.\ref{fig11}(a)-(b), for $\mu=1$ the right boundary can only extract particles from the system and thus activity and current are the same observable. 

With this description, we can compute the average current $\langle q(t)\rangle_{s,h}$ in biased ensembles. In particular, we can recover our previous findings observing the behavior of the current for different values of the biases. For instance, the XX-chain displays an eventual current suppression in ensembles with increasing activities ($s<0$) [see Fig.~\ref{fig1}(a)], while for the diffusive classical SSEP, whose generator is given by \eqref{ME} with bulk field $E=0$, a larger average number of events at the boundary favors an increased net flow of particles [see Fig.~\ref{fig1}(b)].
\begin{figure}[t]
\includegraphics[scale=0.30]{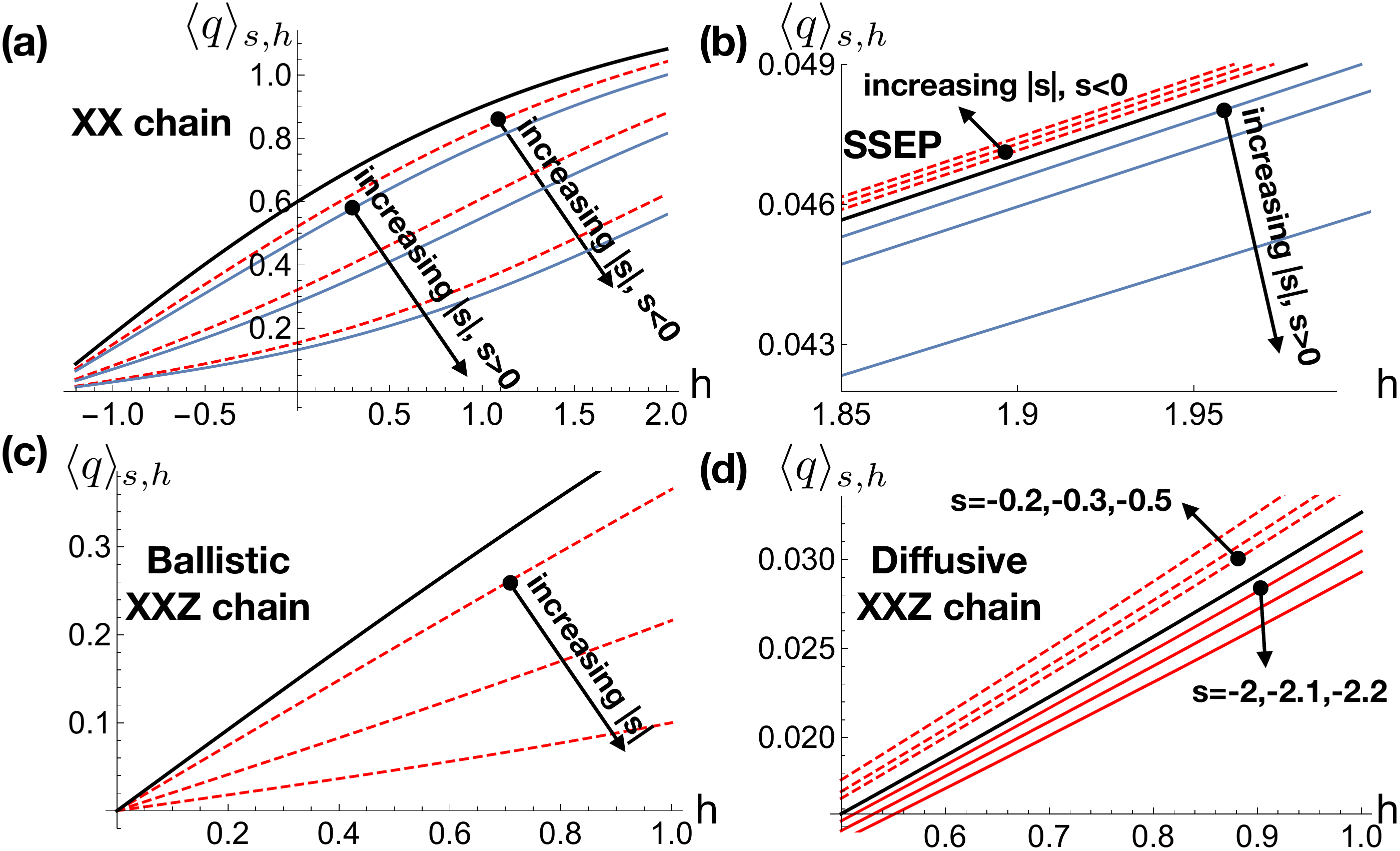}
\caption{\small Current $\langle q(t)\rangle_{s,h}$ computed in ensembles with modified probabilities $\pi_{K,Q}^{s,h}$, favoring or suppressing dynamical realizations according to the value of current and activity. Thick black lines are for typical (non-biased) activity. {\bf Top:} Dashed lines are active ensembles, solid lines inactive ones. {\bf (a)} XX-chain $\mu=0.6$, $|s|=1,2,3$. {\bf (b)} SSEP $L=21$, $|s|=0.5,1,2$, for $\mu=0$. {\bf Bottom:}
{\bf (c)} XXZ-chain, $\delta_z=0.3$, ballistic phase, $\mu=0$. $|s|=1,2,3$, $s>0$ and $s<0$ are indistinghuishable at same $|s|$. {\bf (d)} XXZ-chain, $\delta_z=4$, diffusive phase $\mu=0$. Only active ensembles are displayed: for small $|s|$ currents increase while for large $|s|$ are suppressed.}
\label{fig4}
\end{figure}

Let us now consider the quantum XXZ-chain, whose Hamiltonian is the one of Eq.~\eqref{H} with finite value of $\delta_z$. This system undergoes a phase transition controlled by the anisotropy $\delta_z$ \cite{PhysRevLett.107.137201,Ljubotina:2017aa}, from ballistic ($\delta_z<1$) to diffusive ($\delta_z>1$) transport, and further presents anomalous current fluctuations in the diffusive regime \cite{PhysRevB.90.115156}. It is the simplest, yet non-trivial, model that we can exploit to understand whether the presence of interactions in the Hamiltonian changes the rare behavior of quantum spin chains. For $\mu=0$, we observe, in the ballistic phase, that active and inactive ensembles, characterized by the same $|s|$, show the same suppressed value of the current [see Fig.~\ref{fig4}(c)]. In the diffusive regime, Fig.~\ref{fig4}(d), for small biases towards active realizations, currents tend to increase with the activity, as it happens for classical systems. However, in very active ensembles, there is a departure from the classical diffusive behavior, manifested in the suppression of the current. Since we observe the same for the XX-chain with dephasing at the same $\mu=0$, this particular behavior with $s$ seems to be related to the diffusive regime more than to the presence of interactions in the Hamiltonian. We further considered the classical asymmetric exclusion process with finite field $E$; this model is ballistic \cite{PhysRevLett.95.240601,PhysRevLett.107.010602,PhysRevLett.109.170601}, but, contrary to the quantum ballistic case, larger activities are associated to larger currents [\emph{cf.} Fig.~\ref{fig1}(b)]. 

All these findings lead us to the following conclusion:  in boundary-driven spin chains quantumness manifests in a particular behavior of dynamical fluctuations, which is not dependent on whether the average transport is ballistic or diffusive. The origin of this is the Zeno effect: in very active realizations, sites at the boundaries of the chain are repeatedly disturbed by particle injection or ejection from the reservoir and, thus, the quantum coherent transport is frozen. Via perturbation theory on the tilted operator, we can extend our numerical findings to generic Hamiltonians. Indeed, we find that, with or without bulk dephasing, $\forall\mu\neq1$ and $\forall L$, particle transport is suppressed for ensembles with very large average activity ($s\to-\infty$) \cite{SM},
$$
\lim_{s\to-\infty}\la q(t)\ra_{s,h}=0\, .
$$
As this result is independent of the specific system's Hamiltonian, we have shown that this current suppression is a universal feature of quantum spin chains. Moreover, for $|s|\gg h$, the time-averaged activity $\langle k(t)\rangle_{s,h}\propto \e^{-s}$, and one has 
$
\langle q(t)\rangle_{s,h}\propto \frac{1}{\langle k(t)\rangle_{s,h}}\, ,
$
showing a diffusive-like scaling of the particle transport with the average rate of events at the boudary \cite{SM}.\\

\noindent {\bf \em Spatial structure of trajectories.} Now that we know how large fluctuations allow to discriminate between quantum and classical transport, it is interesting to understand their spatial configuration. A key-quantity capturing relevant features of density correlations is the structure factor \cite{Popkov2011,Jack2015,Karevski2017,PhysRevE.96.052118}
$$
S(p):=\frac{2}{L}\sum_{m,k=1}^{L}\sin(k\,  p)\sin(m\, p)\, C_{m k}(s,h)\, ,
$$
with $p=\frac{\pi}{L}p'$, for $p'=1,2,\dots L-1$, and $C_{m k}(s,h)$ the density-density covariance matrix in the biased $s,h$-ensemble of trajectories. Defining the expectation $\la O\ra_{s,h}=\Tr\left(O\,\ell_{s,h}^{1/2}r_{s,h}\ell_{s,h}^{1/2}\right)$, with $\ell_{s,h},r_{s,h}$ the left, respectively right, eigenmatrix of the tilted-operator $\mathcal{L}_{s,h}$ associated to the eigenvalue $\psi(s,h)$, the density-density correlations can be computed as  \cite{Garrahan2010,PhysRevE.96.052118,carollo2017}
$$
C_{m k}(s,h)=\la n_m n_k\ra_{s,h}-\la n_m\ra_{s,h}\la n_k \ra_{s,h}\, .
$$
For the XX-chain, reconstructing the matrices $r_{s,h},\ell_{s,h}$ from the fermionic formulation is not an easy task. Nonetheless, we can very well approximate these correlations with $C_{m k}(s,h)\approx\la \la n_m n_k\ra\ra_{s,h}-\la \la  n_m\ra\ra_{s,h}\la \la n_k \ra\ra_{s,h}$ \cite{SM}, being $\langle\langle O\rangle\rangle_{s,h}=\Tr\left(\ell_{s,h}\, O \,r_{s,h}\right)$. This functional can be computed in the fermionic language as  $ \langle\langle O\rangle\rangle_{s,h}=\langle {\bf L}_{s,h}| O({\bf a})|{\bf R}_{s,h}\rangle$, with $O({\bf a})$ being the operator $O$ written in terms of Majorana fermions. 

\begin{figure}[t]
\centering
\includegraphics[scale=0.25]{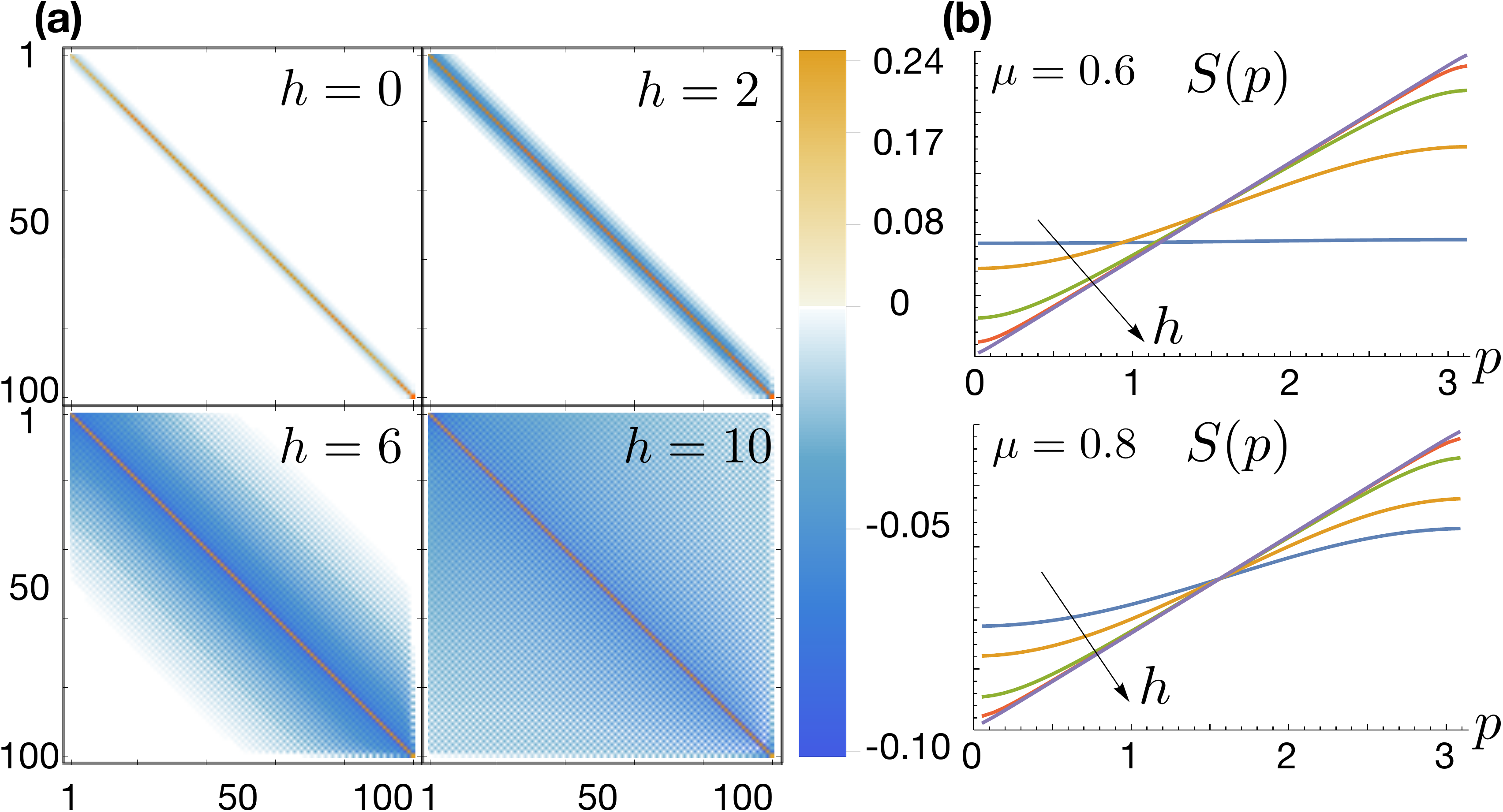}
\caption{\small {\bf (a)} Density correlations $C_{m k}$ for $\mu=0.6$, bias $s=-3$,  $L=100$ and different values of $h$. Favoring realizations with large currents we observe how long-range (anti-)correlations are developed. {\bf (b)} Structure factor $S(p)$ with a linear behavior, for small $p$ in large current events, signalling hyperuniformity. {\bf Top}: $s=-3,\mu=0.6$, and $h=0,3,5,7,10$, $L=100$. {\bf Bottom}: $s=0,\mu=0.8$ and $h=0,1,3,5,7$, $L=50$. }
\label{fig2}
\end{figure}
Density correlations in the typical steady-state dynamics of the XX-chain are extended at most to nearest-neighbors.  
Very different is the behavior for increasing values of the current bias $h$: in these ensembles, characterized by large currents, we observe very weak but longer range correlations spread all along the chain. These have the usual anti-correlated structure suppressing density fluctuations and signalling hyperuniformity \cite{Torquato2016,Karevski2017} (see Fig.~\ref{fig2}).
Conversely, when favoring trajectories with large number of boundary events, density correlations are destroyed. In very active ensembles ($|s|\gg|h|$), the system actually tends to be completely uncorrelated, with an almost flat structure factor. This happens for all considered quantum models, where  large boundary activities break the long-range correlations necessary to sustain efficient particle transport, confirming that the current suppression is caused by the Zeno effect. In stark contrast, in very active realizations of classical chains density correlations are not diminished. 

Also in this case, the XXZ-chain shows an anomalous behavior \cite{PhysRevB.90.115156}. While the other quantum models exhibit density-density anti-correlations and hyperuniformity for large currents, the XXZ-chain displays in the diffusive regime anti-correlations between nearest-neighbors, with all other sites positively correlated. The resulting structure factor does not show an hyperuniform behavior.
\\

\noindent {\bf \em Conclusions.} We conducted  a systematic exploration of large fluctuations in classical and quantum spin chains. Our results here show that the quantum/classical nature of the particle transport is not apparent from the typical behaviour of the dynamics but only becomes apparent after a careful examination of dynamical fluctuations.\\

\begin{acknowledgments}
The research leading to these results has received funding
from the European Research Council under the European
Unions Seventh Framework Programme (FP/2007-
2013)/ERC Grant Agreement No. 335266 (ESCQUMA),
and the EPSRC Grant No. EP/M014266/1. I.L.
gratefully acknowledges funding through the Royal Society
Wolfson Research Merit Award.
\end{acknowledgments}

\bibliographystyle{apsrev4-1}
\bibliography{qhyperuniform}
\onecolumngrid
\newpage

\renewcommand\thesection{S\arabic{section}}
\renewcommand\theequation{S\arabic{equation}}
\renewcommand\thefigure{S\arabic{figure}}
\setcounter{equation}{0}

\begin{center}
{\Large Current fluctuations in boundary-driven quantum spin chains: Supplemental Material}
\end{center}

\subsection{Cumulant generating function of the XX-chain.}
In this section, we will briefly show how to write the tilted-operator for the XX-chain in the third quantization formalism \cite{prosen2008third,prosen2010spectral,Znidaric2014} and compute its eigenvalue with largest real part.  Introducing, via Jordan-Wigner transformation, the Majorana operators for $k=1,2,\dots L$, 
\begin{equation}
w_{2k-1}= \sigma_{\rm x}^{(k)}\prod_{h=1}^{k-1}\sigma_{\rm z}^{(h)},\qquad w_{2k}=\sigma_{\rm y}^{(k)}\prod_{h=1}^{k-1}\sigma_{\rm z}^{(h)}\, ,
\end{equation}
such that $\{w_k,w_h\}=2\delta_{k,h}$, one can write the XX-chain Hamiltonian as 
$$
H=-i\sum_{k=1}^{L-1}\left(w_{2k}w_{2k+1}-w_{2k-1}w_{2k+2}\right)\, .
$$
Also jump operators appearing in the boundary dissipative contribution can be written in the following way:
\begin{equation}
\begin{split}
\sigma_{+}^{(1)}&=\frac{1}{2}\left(w_1+iw_2\right)\, ,\hspace{1.8cm}
\sigma_{-}^{(1)}=\frac{1}{2}\left(w_1-iw_2\right)\, ,\\
\sigma_{+}^{(L)}&=-Z\left(w_{2L-1}+iw_{2L}\right)\, ,\qquad
\sigma_{-}^{(L)}=-Z\left(w_{2L-1}-iw_{2L}\right)\, ,\\
\end{split}
\end{equation}
where $Z$, the parity operator, is $Z=\prod_{k=1}^{L}\sigma_{\rm z}^{(k)}$. 

With the help of the Majorana fermions, we can as well construct a basis for the space of operators. A generic element of such a basis reads
$$
B_{\vec{\alpha}}=\prod_{k=1}^{2L}w_k^{\alpha_k},\qquad  \text{with} \quad \alpha_k=0,1\, .
$$
Through these elements, one can define a vector space formed by the vectors $|B_{\vec{\alpha}}\rangle$, and embedded with the inner product 
$\langle B_{\vec{\beta}}|B_{\vec{\alpha}}\rangle=\frac{1}{2^L}\Tr\left(B_{\vec{\beta}}^\dagger\, B_{\vec{\alpha}}\right)$. Focusing on the even subspace of these operators $B_{\vec{\alpha}}$ (namely the ones for which $\sum_{k}\alpha_k$ is an even number), which is preserved by the action of the tilted operator $\mathcal{L}_{s,h}$, and on which the action of $Z$ is trivial, one can show \cite{prosen2008third,prosen2010spectral,Znidaric2014} that $\mathcal{L}_{s,h}$ can be written as a linear map  
$$
\hat{\mathcal{L}}_{s,h}={\bf \hat{a}}\cdot A\cdot {\bf \hat{a}}-4\, ,
$$ 
acting on the corresponding (even) vector subspace. $A$ is the so-called shape matrix, and the vector ${\bf \hat{a}}$ is a vector of $4L$ new Majorana operators, $\{{\bf \hat{a}}_h,{\bf \hat{a}}_k\}=\delta_{h,k}$, such that 
$$
\sqrt{2}\, {\bf \hat{a}}_{2k-1}|B_{\vec{\alpha}}\rangle=|w_k\, B_{\vec{\alpha}}\rangle\,, \qquad k=1,2,\dots 2L.
$$

The shape matrix $A$, for the considered tilted operator, assumes the following form (coinciding for $s=0$ to what found in \cite{Znidaric2014})
\begin{equation*}
A=\begin{pmatrix}
\hat{B}_{0,0}(-\mu)& \hat{H}&{\bf 0}&\cdots &\cdots & {\bf 0}\\
\hat{H}& {\bf 0} & \hat{H}& \ddots &\ddots & {\bf 0}\\
 {\bf 0} &\hat{H}&  {\bf 0}& \ddots & \ddots & {\bf 0}\\
\vdots& \ddots & \ddots & \ddots & \ddots&\vdots\\
\vdots&  \ddots& \ddots& \ddots  & {\bf 0}& \hat{H}\\
 {\bf 0}&  {\bf 0} & {\bf 0}&  \cdots &  \hat{H} &\hat{B}_{s,h}(\mu)\\
\end{pmatrix}\, ,
\end{equation*}
with $\hat{H}=i\sigma_{\rm y}\otimes {\bf 1}_2$, and 
$$
\hat{B}_{s,h}(\mu)=f^1_{s,h}(\mu)\, \sigma_{\rm y}\otimes\sigma_{\rm z}+f^2_{s,h}(\mu)\, \sigma_{\rm y}\otimes \sigma_{\rm x}+f^3_{s,h}(\mu)\, {\bf 1}_{2}\otimes \sigma_{\rm y}\, ,
$$
together with 
\begin{equation}
\begin{split}
f^1_{s,h}(\mu)&=-\mu\, ,\\
f^2_{s,h}(\mu)&=-i\e^{-s}\left(\mu\cosh(h)+\sinh(h)\right)\, ,\\
f^3_{s,h}(\mu)&=-\e^{-s}\left(\cosh(h)+\mu\sinh(h)\right)\, .
\end{split}
\end{equation}

This matrix can be written in a tensor product form 
\begin{equation*}
\begin{split}
A=&D_1\otimes \left(f^1_{0,0}(-\mu)\, \sigma_{\rm y}\otimes\sigma_{\rm z}+f^2_{0,0}(-\mu)\, \sigma_{\rm y}\otimes \sigma_{\rm x}+f^3_{0,0}(-\mu)\, {\bf 1}_{2}\otimes \sigma_{\rm y}\right)+\\+&\tilde{H}\otimes \hat{H}
+D_L\otimes \left(f^1_{s,h}(\mu)\, \sigma_{\rm y}\otimes\sigma_{\rm z}+f^2_{s,h}(\mu)\, \sigma_{\rm y}\otimes \sigma_{\rm x}+f^3_{s,h}(\mu)\, {\bf 1}_{2}\otimes \sigma_{\rm y}\right)\, ,
\end{split}
\end{equation*}
where $(D_N)_{m,k}=\delta_{k,N}\delta_{k,m}$ and with $\tilde{H}$ being the $L\times L$ matrix whose non-zero elements are only $\tilde{H}_{k,k+1}=\tilde{H}_{k-1,k}=1$. All terms have, as second entry of the tensor product, either an identity or a  $\sigma_{\rm y}$. It proves therefore convenient to reshape the above matrix moving the second and the third entries of the tensor product to the first, respectively second position. In this new representation the matrix reads
\begin{equation*}
\begin{split}
A'=&\left(f^1_{0,0}(-\mu)\, \sigma_{\rm y}\otimes\sigma_{\rm z}+f^2_{0,0}(-\mu)\, \sigma_{\rm y}\otimes \sigma_{\rm x}+f^3_{0,0}(-\mu)\, {\bf 1}_{2}\otimes \sigma_{\rm y}\right)\otimes D_1+\\+&i\sigma_{\rm y}\otimes{\bf 1}_2\otimes\tilde{H}
+ \left(f^1_{s,h}(\mu)\, \sigma_{\rm y}\otimes\sigma_{\rm z}+f^2_{s,h}(\mu)\, \sigma_{\rm y}\otimes \sigma_{\rm x}+f^3_{s,h}(\mu)\, {\bf 1}_{2}\otimes \sigma_{\rm y}\right)\otimes D_L\, .
\end{split}
\end{equation*}
Now, we apply a rotation on the first term of the tensor product, bringing $\sigma_{\rm y}$ to its diagonal form ($U\sigma_{\rm y}U^\dagger=\sigma_{\rm z}$, with $U$ as in the main text), obtaining 
\begin{equation*}
\begin{split}
A''={\bf 1}_2\otimes\sigma_{\rm y}\otimes\left(f^3_{0,0}(-\mu)D_1+f^3_{s,h}(\mu)D_L\right)+&\sigma_{\rm z}\otimes \Bigg(f_{0,0}^1(-\mu)\sigma_{\rm z}\otimes D_1+f_{0,0}^2(-\mu)\sigma_{\rm x}\otimes D_1+\\
+&i{\bf 1}_2\otimes \tilde{H}+f_{s,h}^1(\mu)\sigma_{\rm z}\otimes D_L+f_{s,h}^2(\mu)\sigma_{\rm x}\otimes D_L\Bigg)\, .
\end{split}
\end{equation*}
We can then collect terms introducing a matrix $X$, so that one has 
$$A''=\begin{pmatrix}
X&0\\
0&-X^T
\end{pmatrix}\, ;
$$
the matrix $X$ is given by 
$X={\bf 1}_2\otimes i\tilde{H}+\Gamma_1+\Gamma_L$, with $\Gamma_1=B_{0,0}(-\mu)\otimes D_1$ and $\Gamma_L=B_{s,h}(\mu)\otimes D_L$, where 
$$
B_{s,h}(\mu)=\left(
\begin{array}{cc}
 -\mu  & -i e^{-(s+h)} (\mu -1) \\
 -i e^{-(s-h)} (\mu +1) & \mu  \\
\end{array}
\right)\, .
$$

The reordering that we performed on the tensor product affects also the vector of Majorana fermions ${\bf \hat{a}}$. This can be accounted for by introducing a new vector ${\bf a}$, made as follows
$$
{\bf a}=\Big({\bf \hat{a}}_1,{\bf \hat{a}}_5,{\bf \hat{a}}_{9},\dots {\bf \hat{a}}_{1+4(L-1)},{\bf \hat{a}}_2,{\bf \hat{a}}_6,{\bf \hat{a}}_{10},\dots {\bf \hat{a}}_{2+4(L-1)},{\bf \hat{a}}_3,{\bf \hat{a}}_7,{\bf \hat{a}}_{11},\dots {\bf \hat{a}}_{3+4(L-1)},{\bf \hat{a}}_4,{\bf \hat{a}}_8,{\bf \hat{a}}_{12},\dots {\bf \hat{a}}_{4L}\Big)^T\, .
$$ 
This shows that the generator can be written as in equation \eqref{GenTQ} of the main text:
$$
\hat{\mathcal{L}}_{s,h}={\bf a}\cdot U^\dagger \begin{pmatrix}
X&0\\
0&-X^T
\end{pmatrix}
 U\cdot{\bf a}-4\, .
$$
We now want to find the largest real eigenvalue of the above generator. Assuming $X$ to be diagonalizable, there exists a matrix $P$, such that $X=P\Lambda P^{-1}$, with $\Lambda$ diagonal; thus
$$
U^\dagger \begin{pmatrix}
X&0\\
0&-X^T
\end{pmatrix}U=U^\dagger \begin{pmatrix}
P&0\\
0&P^{-T}
\end{pmatrix}\begin{pmatrix}
\Lambda&0\\
0&-\Lambda
\end{pmatrix}\begin{pmatrix}
P^{-1}&0\\
0&P^T
\end{pmatrix} U\, .
$$
Defining $V=\begin{pmatrix}
P^{-1}&0\\
0&P^T
\end{pmatrix} U$, with analogous calculation to those of Ref.~\cite{prosen2010spectral}, one finds
$$
\hat{\mathcal{L}}_{s,h}={\bf a}\cdot V^T\begin{pmatrix}
0&-\Lambda\\
\Lambda&0
\end{pmatrix}
V \cdot{\bf a}-4\, .
$$
The matrix $V$ implements a generalised rotation acting on the vector ${\bf a}$, in such a way that it introduces $4L$ almost canonical fermionic creation and annihilation operators,  
$$
\begin{pmatrix}
b\\
b'
\end{pmatrix}=V\cdot {\bf a}\, ,
$$
obeying $\left\{b_h,b'_k\right\}=\delta_{h,k}$, and with all other anticommutation relations being zero. These creation and annihilation operators are the normal master modes of the tilted operator. With these, one finds
$$
\hat{\mathcal{L}}_{s,h}=\sum_{j=1}^{2L}\Lambda_j\left(b'_jb_j-b_jb'_j\right)-4\, ,
$$
and using the anticommutation relations
$$
\hat{\mathcal{L}}_{s,h}=2\sum_{j=1}^{2L}\Lambda_jb'_jb_j-4-\sum_{j=1}^{2L}\Lambda_j=2\sum_{j=1}^{2L}\Lambda_jb'_jb_j-4\, ,
$$
where the last equality comes from the fact that $X$ is traceless. The dependence on the biases $s,h$ and on the parameter $\mu$ is encoded in the eigenvalues $\Lambda_j$ of the matrix $X$, as well as in the rotation matrix $V$. Introducing right and left vacuum, $|0\rangle$,$\langle \tilde{0}|$, which are annihilated by $b_{m}$, and $b'_m$ respectively, one finds that the eigenvalue with the largest real part corresponds to the right eigenvector $|{\bf R}_{s,h}\rangle=\prod_{m\in\Lambda^{+}} b'_{m}|0\rangle$, as well as to the left one $\langle{\bf L}_{s,h}|=\langle\tilde{0}|\prod_{m\in\Lambda^{+}} b_{m}$, and is given by 
$$
\psi(s,h)=2\sum_{m\in\Lambda^+}{\rm Re}(\Lambda_m)-4\, ,
$$
with $\Lambda^+$ being the set of $m$ for which ${\rm Re}(\Lambda_m)>0$.

\subsection{Perturbation theory on the tilted-operator.}
Let us start by writing explicitly all terms of the tilted operator $\mathcal{L}_{s,h}$ with a generic Hamiltonian $H$:
\begin{equation}
\begin{split}
\mathcal{L}_{s,h}[\rho]&=-i[H,\rho]+\gamma_0\sigma_+^{(1)}\rho\sigma_-^{(1)}-\frac{\gamma_0}{2}\left\{\rho,\sigma_-^{(1)}\sigma_+^{(1)}\right\}+\gamma_1\sigma_-^{(1)}\rho\sigma_+^{(1)}-\frac{\gamma_1}{2}\left\{\rho,\sigma_+^{(1)}\sigma_-^{(1)}\right\}+\\
&+\e^{-s}\Bigg[\gamma_1\e^{-h}\sigma_+^{(L)}\rho\sigma_-^{(L)}+\gamma_0\e^{h}\sigma_-^{(L)}\rho\sigma_+^{(L)}\Bigg]-\frac{\gamma_1}{2}\left\{\rho,\sigma_-^{(L)}\sigma_+^{(L)}\right\}-\frac{\gamma_0}{2}\left\{\rho,\sigma_+^{(L)}\sigma_-^{(L)}\right\}\, .
\end{split}
\label{explicit}
\end{equation}
For large negative $s$ we see that there is a part of the above map which is predominant. Defining 
$$
\mathcal{K}[\rho]=\Bigg[\gamma_1\e^{-h}\sigma_+^{(L)}\rho\sigma_-^{(L)}+\gamma_0\e^{h}\sigma_-^{(L)}\rho\sigma_+^{(L)}\Bigg]\, ,
$$
and collecting in $\mathcal{W}[\rho]$ all remaining terms appearing on the right-hand side of equation \eqref{explicit}, we can write the tilted operator as
$$
\mathcal{L}_{s,h}[\rho]=\e^{-s}\mathcal{K}[\rho]+\mathcal{W}[\rho]\, .
$$
When considering large negative $s$, $\mathcal{L}_{s,h}=\e^{|s|}\left(\mathcal{K}+\e^{-|s|}\mathcal{W}\right)$, with $\e^{-|s|}$ a small number, showing that we can apply perturbation theory in order to consider the correction to the dominant term $\mathcal{K}$ due to the map $\mathcal{W}$. To proceed, one needs first to diagonalize the map $\mathcal{K}$. This acts in a non-trivial way only on the last site of the chain; we therefore consider operators of the form $x\otimes y$,
where $x$ is an operator acting on the first $L-1$ sites of the chain, while $y$ acts only on the last one. We thus have
$$
\mathcal{K}[x\otimes y]=x\otimes \hat{\mathcal{K}}[y]\, ,\qquad  \text{with} \qquad \hat{\mathcal{K}}[y]=\Bigg[\gamma_1\e^{-h}\sigma_+y\sigma_-+\gamma_0\e^{h}\sigma_-y\sigma_+\Bigg]\, .
$$
It can be shown that the largest eigenvalue of the map $\hat{\mathcal{K}}$ is given by $\sqrt{\gamma_0\gamma_1}$, with associated right eigenmatrix $r$ and left one $\ell$ being as follows
$$
r=\begin{pmatrix}
\frac{\sqrt{\gamma_1}}{\sqrt{\gamma_1}+\e^{-h}\sqrt{\gamma0}}&0\\
0&\frac{\sqrt{\gamma_0}}{\sqrt{\gamma_0}+\e^{h}\sqrt{\gamma_1}}
\end{pmatrix}, \qquad \qquad \ell=\frac{1}{2}\begin{pmatrix}
\frac{\sqrt{\gamma_1}+\e^{-h}\sqrt{\gamma0}}{\sqrt{\gamma_1}}&0\\
0&\frac{\sqrt{\gamma_0}+\e^{h}\sqrt{\gamma_1}}{\sqrt{\gamma_0}}
\end{pmatrix}.
$$
Given any operator $x$ acting on the first $L-1$ sites of the chain, one has $\mathcal{K}[x\otimes r]=\sqrt{\gamma_0\gamma_1} x\otimes r$. This shows that the eigenvalue $\sqrt{\gamma_0\gamma_1}$ is highly degenerate. Taking into account this degeneracy when applying perturbation theory, one has that the eigenvalue with the largest real part of the tilted operator, which is the cumulant generating function $\psi(s,h)$, is given by 
\begin{equation}
\psi(s,h)=\sqrt{\gamma_0\gamma_1}\e^{|s|}+w(h)+O(\e^{-|s|})\, ,
\label{pert}
\end{equation}
where $w(h)$ is the (possibly $h$ dependent) eigenvalue with the largest real part of the  matrix 
$$
w_{ij}=\Tr\left(b^\dagger_i\otimes \ell \,\mathcal{W}[b_j\otimes r]\right)\, ,
$$
with $b_i$ being an element of an operator basis for the first $L-1$ sites of the chain, such that $ \Tr\left(b_i^\dagger b_j\right)=\delta_{i,j}$. Because of the fact that $\ell r=r\ell={\bf 1}_2/2$, and because of the shape of the map $\mathcal{W}$, it is straightforward to show that the matrix $[w_{ij}]$ actually does not depend on $h$ and nor does its largest eigenvalue $w(h)$. The same holds true even if one considers the presence of an extra dephasing term in the Lindblad generator given by
$$
\mathcal{L}^D[\rho]=\gamma_D\sum_{m=1}^{L}\left(n^{(m)}\rho\, n^{(m)} -\frac{1}{2}\left\{\rho, n^{(m)}\right\}\right)\, .
$$
Since $w(h)$ does not depend on $h$, for large biases towards active realizations ($-s\gg1$), the current, given by the first derivative with respect to $h$ of $\psi(s,h)$ [{\emph{cf.} Eq.~\eqref{pert}], is of order  $\la q(t)\ra_{s,h}=\partial_h\psi(s,h)\sim O(\e^{-|s|})$. 

Given also that  
$
\la k(t)\ra_{s,h}=-\partial_s \psi(s,h)\approx e^{|s|}
$,
one recovers, for large negative $s$, the diffusive-like scaling of the current with the average activity,  $\la q(t)\ra_{s,h}\propto \frac{1}{\la k(t)\ra_{s,h}}$. 

\subsection{Approximation of density-density correlations for the computation of the structure factor.}
We show here by numerical evidence that, concerning the computation of the structure factor for the XX-chain, we can approximate the expectation $\langle O\rangle_{s,h}=\Tr\left(O\, \ell_{s,h}^{1/2}r_{s,h}\ell_{s,h}^{1/2}\right)$ with the functional $\langle\langle  O\rangle \rangle_{s,h}=\Tr\left(\ell_{s,h}\, O\, r_{s,h}\right)$.

We verified this for systems with up to $L=10$ sites, always obtaining a satisfactory agreement between the structure factor $S(p)$ computed via the exact correlations $\la n_m n_k\ra_{s,h}-\la n_m\ra_{s,h}\la n_k \ra_{s,h}$ and the approximated structure factor $S_A(p)$ computed with the correlations $\la\la n_m n_k\ra\ra_{s,h}-\la\la n_m\ra\ra_{s,h}\la\la n_k \ra\ra_{s,h}$.  Indeed, as it is possible to appreciate from Fig.~\ref{SFCheck}, there is a very nice agreement between the two. Moreover, from numerical results, we  observed that the error made in computing density-density correlations between bulk sites with the approximated functional is smaller than the one made computing density correlations for sites next to the boundaries. Because of this, we expect the agreement between $S(p)$ and $S_A(p)$ to persist also for larger $L$, as bulk contributions become predominant.
\begin{figure}[h!]
\centering
\includegraphics[scale=0.75]{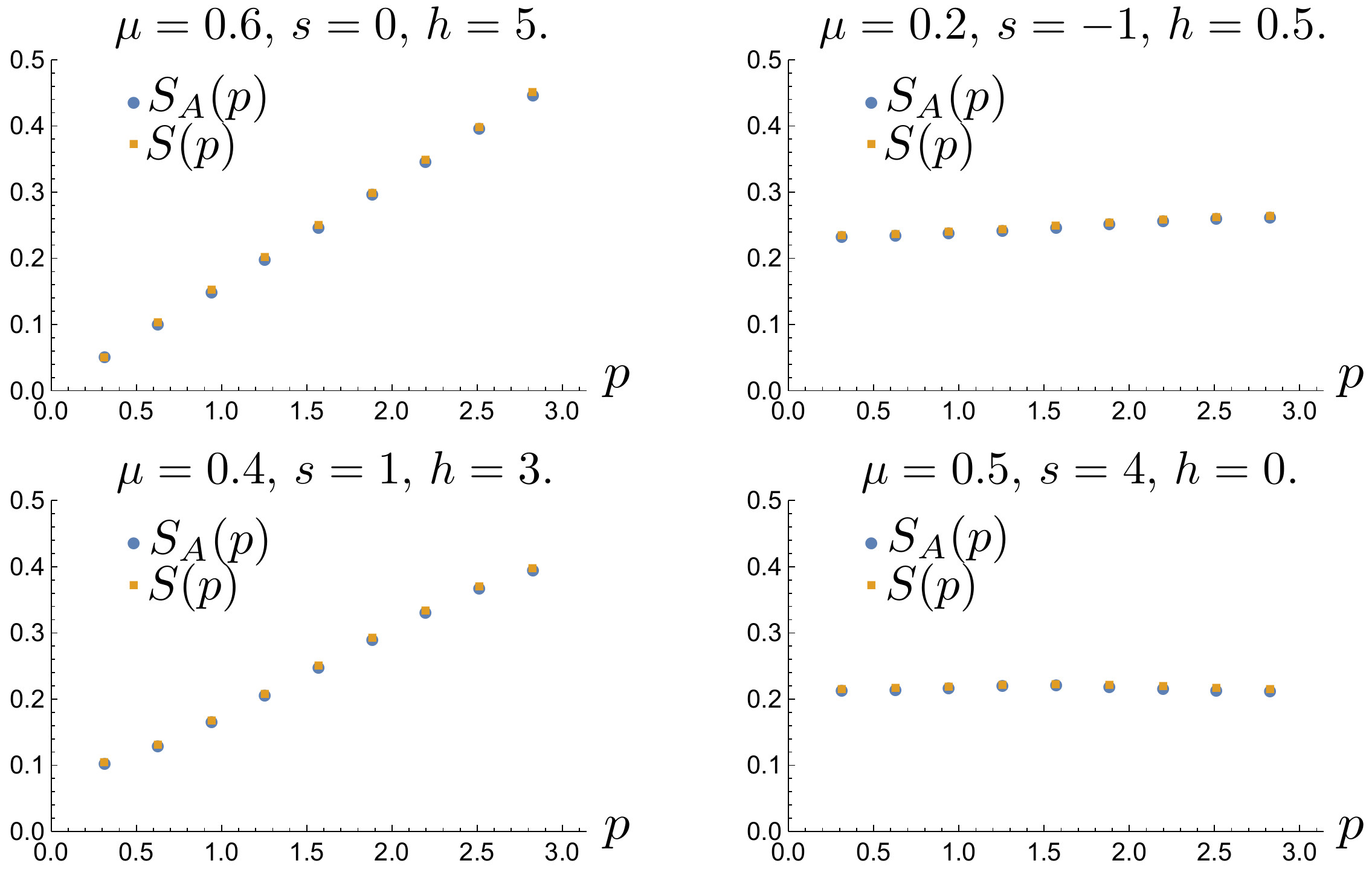}
\caption{\small Comparison between the exact structure factor $S(p)$ and the approximated one $S_A(p)$ for an XX-chain with $L=10$ and different combinations of the various parameters. The nice agreement between the two quantities shows that one can rely on this approximation to gain insight on the structure of the various ensembles.}
\label{SFCheck}
\end{figure}
\newpage

\end{document}